# Use of the geometric mean as a statistic for the scale of the coupled Gaussian distributions

Kenric P. Nelson[1], Mark A. Kon[1] and Sabir R. Umarov[2]

**Highlights:**

1. A functional relationship is established between the scale of a coupled Gaussian distribution and the geometric mean of the distribution.

2. Numerical evidence is provided suggesting that an estimator of the scale based on this function is unbiased with diminishing variance as the sample size grows.

3. A relationship between the generalized mean and the scale of the coupled Gaussian is also established, but this relationship does not lead to an effective estimator.

4. Fluctuations due to global correlation are shown to be proportional to the nonlinear statistical coupling.

**Abstract:** *The geometric mean is shown to be an appropriate statistic for the scale of a heavy-tailed coupled Gaussian distribution or equivalently the Student's t distribution. The coupled Gaussian is a member of a family of distributions parameterized by the nonlinear statistical coupling which is the reciprocal of the degree of freedom and is proportional to fluctuations in the inverse scale of the Gaussian. Existing estimators of the scale of the coupled Gaussian have relied on estimates of the full distribution, and they suffer from problems related to outliers in heavy-tailed distributions. In this paper, the scale of a coupled Gaussian is proven to be equal to the product of the generalized mean and the square root of the coupling. From our numerical computations of the scales of coupled Gaussians using the generalized mean of random samples, it is indicated that only samples from a Cauchy distribution (with coupling parameter one) form an unbiased estimate with diminishing variance for large samples. Nevertheless, we also prove that the scale is a function of the geometric mean, the coupling term and a harmonic number. Numerical experiments show that this estimator is unbiased with diminishing variance for large samples for a broad range of coupling values.*

## 1. Introduction

The estimation of uncertainty rests principally on the variance and entropy of a random variable, but in nonlinear systems a broader perspective on statistical analysis is required. One manifestation of the complexity of nonlinear systems is the role of power law statistics, which are not characterized by the variance of related random variables. Renyi, Tsallis and other investigators have shown that generalizations of the entropy function can be used to explain the role of power law statistics in complex systems [1–4]. It is not as widely appreciated, that both the Renyi and Tsallis entropies of a distribution utilize the generalized mean to combine the probabilities of a distribution. The Rényi entropy translates the generalized mean of the distribution to an entropy scale using the natural logarithm, while the Tsallis entropy utilizes a generalized logarithm [5,6]. The special case of Boltzmann-Gibbs-Shannon entropy is a translation of the geometric mean by the logarithm. As such the generalized mean is an important statistic for characterizing uncertainty in data arising from complex systems. In this paper we explore the properties of the generalized mean as an estimator of the scale for a family of distributions which have power-law decay.

Two fundamental criteria for estimation are efficiency [7] and sufficiency [8]. An estimator is said to be efficient if it minimizes a loss function, such as the mean-square error. The mean square error of an estimator $\hat{\theta}$ is the sum of the variance and the square of the bias

[1] Boston University; Correspondence: kenricpn@bu.edu, 781-645-8564
[2] University of New Haven

$$MSE = E\left[\left(\hat{\theta}-\theta\right)^2\right] = E\left[\left(\hat{\theta}-E(\hat{\theta})+E(\hat{\theta})-\theta\right)^2\right] = Var(\hat{\theta}) + \left(b(\hat{\theta})\right)^2. \qquad (1)$$

An estimator which is unbiased and minimizes the variance, would thus be an efficient unbiased estimator based on the mean-square error criteria. A stronger criterion is sufficiency, which requires that a statistic utilize all the possible information in a data set with regard to estimating a parameter. More precisely, sufficiency requires that the conditional probability given a statistic is independent of the parameter

$$p(\mathbf{X}=\mathbf{x}|T(\mathbf{X})=t,\theta) = p(\mathbf{X}=\mathbf{x}|T(\mathbf{X})=t). \qquad (2)$$

In other words, given the statistic $t$ of the dataset **x**, knowledge of the parameter does not reduce the uncertainty. A statistic is sufficient if and only if the probability of the dataset conditioned on the parameter can be factored into two functions, one dependent on the statistic and one independent of the statistic,

$$f(\mathbf{x}|\theta) = g(T(\mathbf{x})|\theta)h(\mathbf{x}). \qquad (3)$$

The exponential family is defined in terms of this factorization principal [9]. Given the natural parameters $\eta(\theta)$, their sufficient statistics $T(x)$ and the log partition function $A(\theta)$ the exponential family of distributions is defined as

$$f(x|\theta) = h(x)\exp\big(\eta(\theta)T(x) - A(\theta)\big). \qquad (4)$$

In this paper, we provide numerical evidence that a family of heavy-tail distributions referred to as coupled-exponentials [10], may have an efficient estimator for their scale parameter. If so, there may be a generalization of the factorization theorem for sufficiency. The coupled exponentials are non-exponential except in the degenerate case that the coupling is zero. The coupling refers to a nonlinear deformation of the exponential and logarithmic functions. This generalization is reviewed in Section 2. In Section 3, numerical experiments using the generalized mean as a statistic to estimate the scale parameter of the coupled exponential distributions are explored. Evidence is provided that the geometric mean in particular has potential to be a statistic from which an unbiased efficient estimator of the scale can be formed. Section 4 provides examples of estimating the scale and tail decay of the coupled exponential distributions. Section 5 provides a conclusion and discussion of the results.

## 2. Review of the coupled exponential family of distributions

The investigation focuses on a family of distributions [11] related to the power function $(1+\kappa x)^{\frac{1}{\kappa}}$, which reduces to the exponential function as $\kappa \to 0$. For $\kappa \neq 0$ this function approximates the exponential near $x=0$ and has power-law properties as $x \to \infty$. This function solves the nonlinear differential equation $dy/dx = y^{1-\kappa}$. For this and further reasons (explained below) $\kappa$ is referred to as the *nonlinear statistical coupling* [12] or simply the *coupling*. We use a shorthand notation for the power function emphasizing its role as a deformation of the exponential function [13] by writing

$$\exp_\kappa^a(x) \equiv (1+\kappa x)_+^{\frac{a}{\kappa}}, \tag{5}$$

where the parameter measures the level of deformation from the exponential function, while the exponent $a$ is simply a power parameter; here we have defined $(y)_+ = \max(0,y)$. Given an argument $\left|\frac{x-\mu}{\sigma}\right|^\alpha$ with mean $\mu$, scale $\sigma$ and power $\alpha$, a family of distributions is defined which we will refer to as the *coupled exponential family*.

*Definition 1*: A coupled exponential family of distributions is defined by

$$f(x;\mu,\sigma,\kappa,\alpha) \equiv \frac{1}{Z(\sigma,\kappa,\alpha)} \exp_\kappa^{\frac{1+\kappa}{\alpha}} \left(\frac{|x-\mu|^\alpha}{\sigma^\alpha}\right) = \left( Z(\sigma,\kappa,\alpha) \left( 1+\kappa \frac{|x-\mu|^\alpha}{\sigma^\alpha} \right)_+^{\frac{1+\kappa}{\alpha\kappa}} \right)^{-1},$$

$$Z(\sigma,\kappa,\alpha) = \begin{cases} \frac{2\sigma}{\alpha} \kappa^{-\frac{1}{\alpha}} B\left(\frac{1}{\alpha\kappa},\frac{1}{\alpha}\right), & \kappa > 0 \\ \frac{2\sigma}{\Gamma\left(1+\frac{1}{\alpha}\right)}, & \kappa = 0 \\ \frac{2\sigma}{\alpha}(-\kappa)^{-\frac{1}{\alpha}} B\left(1-\frac{1+\kappa}{\alpha\kappa},\frac{1}{\alpha}\right) & -1 < \kappa < 0 \end{cases} \tag{6}$$

$$\sigma \geq 0, 0 < \alpha \leq 3,$$

where $\Gamma$ and $B$ are Euler's Gamma and Beta function respectively, and $Z$ is the normalization. The exponent term $1+\kappa$ arises from the dimensions of the distribution. The cumulative distribution has an asymptotic power of $-\frac{1}{\kappa}$ while the asymptotic power of the density is decremented by one for each dimension. Thus for a *d*-dimensional distribution, the exponent term would include the factor $1+d\kappa$.

Two important members of this family are $\alpha=1$, which is the generalized Pareto distribution and $\alpha=2$, which is the Student's t distribution. These distributions have been derived and utilized in a variety of contexts. The generalized Pareto distribution, referred to here as the coupled exponential distribution where $\kappa$ is the shape parameter, is often utilized to model heavy-tail distributions above a threshold $\mu$ [14]. The Student's t distribution, referred to here as the coupled Gaussian distribution where $\kappa$ is the inverse of the degree of freedom, arises from estimation of the variance of a Gaussian random variable given limited samples [15]. Another origin of a coupled Gaussian random variable is as the division of a Gaussian random variable by a Chi-squared random variable [8]. Our interest will be in estimation of the scale $\sigma$ when the coupling $\kappa$ is positive which is the domain of heavy-tail distributions. When $\kappa=0$ the distribution is the generalized Gaussian which is part of exponential family. For negative values of $\kappa$ the distributions have compact-support for which the scale can be estimated using the standard deviation.

A more recent interest in these distributions has arisen in the statistical modeling of complex systems which are influenced by nonlinear dynamics. In these systems the nonlinear dynamics has the effect of coupling the statistical states [6,12] over long spatial and/or temporal ranges resulting in the information of system not having the typical property of extensivity and/or additivity. Investigators [2,11,16–21] have shown that the coupled exponential family of distributions maximize a generalized entropy function and are attractors in the generalized central limit theorem [22,23]. Tsallis [24] investigated the properties of an escort distribution $f^{(q)}(x) \equiv f^q(x) / \int_{x \in X} f^q(x)$, with $q = 1 + \frac{\alpha \kappa}{1 + d\kappa}$ as an alternative constraint for determining a distribution which maximizes a generalized entropy There is substantial literature seeking to interpret the parameter $q$ in terms of the properties of complex systems [25–28], though there is no consensus. In the case of $f^q(x)$, $q$ is the "fractional" number of independent random variables $X$ and $f^q(x)$ is the distribution of $q$ independent occurrences $x$ of $X$. This approach has been shown to model non-extensivity, long-range correlation and other properties of complex systems. Nevertheless, the degree of non-extensivity or long-range correlation is not itself defined or quantified by $q$. A candidate was proposed in [12] using the relationship $r = 1 - q$. A measure of risk bias related to this term has subsequently been derived for in [29]; nevertheless, examination of multivariate systems in [10] shows that separation of the dimensions and power of the state variable is necessary if the global correlation is to be quantified.

One approach to modeling a system with global correlation is to consider a sequence of exchangeable random variables [30]. Exchangeable variables generalize the concept of independent identically distributed (iid) random variables to dependent sequences in which the marginal distributions are identical and the joint distribution is symmetric under permutation, i.e. the order of the variables can be exchanged. An example is the generalized Gaussian distribution with random inverse scales having a gamma distribution. The result is exchangeable random variables distributed as the coupled exponential. Unfortunately, because global correlation can take many different forms, there is not a single metric which is universally accepted as the measure of this property. The difference between the mutual information of the product of the marginal distributions (under independence) and the joint distribution is perhaps the most well studied. Here we focus on the degree of mixing measured by the relative variance of the fluctuations. While not the correlation itself, this measures one of the physical mechanisms creating global correlation.

The relative variance of a random variable $X$ is

$$\text{RelVar}(X) \equiv \frac{\text{Var}(X)}{\text{Mean}^2(X)} = \frac{E(X^2) - E^2(X)}{E^2(X)}, \quad (7)$$

where $E(X)$ is the expected value or mean. If the inverse scale $\theta^{-\alpha}$ of the generalized Gaussian distribution is itself a fluctuating random variable distributed as a gamma distribution, the resulting distribution is the coupled exponential family,

$$\frac{1}{Z(\sigma,\kappa,\alpha)}\left(1+\kappa\frac{|x-\mu|^{\alpha}}{\sigma^{\alpha}}\right)^{\frac{1+\kappa}{-\alpha\kappa}}=\int_{0}^{\infty}\frac{1}{Z(\theta,0,\alpha)}\exp\left(\frac{|x-\mu|^{\alpha}}{\alpha\theta^{\alpha}}\right)g(\theta^{-\alpha})\alpha\theta^{-\alpha-1}d\theta$$

$$g\left(\theta^{-\alpha};\frac{1}{\alpha\kappa},\alpha\kappa\sigma^{-\alpha}\right)=\frac{\left(\alpha\kappa\sigma^{-\alpha}\right)^{\frac{-1}{\alpha\kappa}}}{\Gamma\left(\frac{1}{\alpha\kappa}\right)}\theta^{\alpha-\frac{1}{\kappa}}\exp\left(-\frac{\sigma^{\alpha}}{\alpha\kappa}\theta^{-\alpha}\right).$$

(8)

This gamma distribution has a mean of $\sigma^{-\alpha}$ and a variance of $\alpha\kappa\sigma^{-2\alpha}$, which results in a relative variance for the fluctuating inverse scale of

$$\mathrm{RelVar}\left[\theta^{-\alpha}\right]=\frac{\mathrm{Var}\left[\theta^{-\alpha}\right]}{E^{2}\left[\theta^{-\alpha}\right]}=\alpha\kappa.$$

(9)

Thus the degree of mixing is the product of the two terms creating nonlinearity. $\alpha$ defines the power of the unfluctuating state space. $\kappa$ defines the deviation from the linear differential equations resulting in fluctuations which create global correlation. Rearranging terms gives a physical definition for the nonlinear statistical coupling.

*Definition* 2: The *nonlinear statistical coupling* $\kappa$ is the relative variance of the fluctuations of the inverse scale $\theta^{-\alpha}$ of a generalized Gaussian distribution divided by the power term $\alpha$. The fluctuations are assumed to be distributed by a gamma distribution. From equations (8) and (9) one has

$$\kappa\equiv\frac{1}{\alpha}\mathrm{RelVar}\left[\theta^{-\alpha}\right],$$

(10)

where $\theta^{-\alpha}$ is distributed as a gamma distribution defined in equation (8).

*Remark*: This definition provides evidence that the nonlinear statistical coupling may quantify the physical properties leading to global correlation. The suggestion that the model relates to coupling of the states arose from the fact that the related variable $r=\frac{-\alpha\kappa}{1+\kappa}=1-q$ can be used to rearrange the escort probability as the probability of the all the states coupled together. For discrete variables this relationship is $p_{i}^{(r)}\equiv\frac{p_{i}\prod_{j=1,j\neq i}^{n}p_{j}^{r}}{\sum_{k=1}^{n}p_{k}\prod_{j=1,j\neq k}^{n}p_{j}^{r}}$, which can be interpreted as a nonlinear coupling between states of a globally correlated system.

Since the scale of the coupled exponentials results from fluctuations of the scale of the generalized Gaussian, the standard deviation cannot be used to measure this property. Instead, the scale is determined using the escort probability

$$\sigma = \left( \frac{\int_{x \in X} x^{\alpha} f^{q}(x;\mu,\sigma,\kappa,\alpha) dx}{\int_{x \in X} f^{q}(x;\mu,\sigma,\kappa,\alpha) dx} \right)^{\frac{1}{\alpha}}, \quad q = 1 + \frac{\alpha \kappa}{1+\kappa}. \tag{11}$$

Equation (11) with fixed $\sigma$ is used as a constraint on the maximization of the generalized entropy. The generalized entropy is formulated in terms of the coupled logarithm which is the inverse of the coupled exponential function $\ln_{\kappa} x \equiv \frac{1}{\kappa}(x^{\kappa} - 1); x > 0$. The generalized logarithm has the property $\ln_{\kappa} x^a = a \ln_{\alpha \kappa} x$. Not including the normalizing term Z and the exponent $-1/\alpha$, the inverse of (5) is used to define a generalized entropy function

$$\ln_{\kappa} x^{\frac{-\alpha}{1+\kappa}} = \left( \frac{1}{\kappa} \left( x^{\frac{-\alpha \kappa}{1+\kappa}} - 1 \right) \right); \quad x > 0. \tag{12}$$

In [6] the authors showed that the generalized entropy can be expressed as the generalized logarithm of an *average density* which is determined by the weighted generalized mean of the density. This function defines the *coupled-entropy* $S_{\alpha,\kappa}$.

*Definition 2*: The Coupled Entropy and the Average Density

$$S_{\alpha,\kappa}(f) \equiv -\ln_{\kappa} f_{avg}^{\frac{-\alpha}{1+\kappa}}(x) = -\ln_{\kappa} \left( \int_{x \in X} f^{1+\frac{\alpha \kappa}{1+\kappa}}(x) dx \right)^{\frac{-1}{\kappa}} = \frac{1}{\kappa} \left( 1 - \left( \int_{x \in X} f^{1+\frac{\alpha \kappa}{1+\kappa}}(x) dx \right)^{-1} \right)$$

$$f_{avg}(x) \equiv \left( \int_{x \in X} f(x) f^{\frac{\alpha \kappa}{1+\kappa}}(x) dx \right)^{\frac{1+\kappa}{\alpha \kappa}} = \left( \int_{x \in X} f^{1+\frac{\alpha \kappa}{1+\kappa}}(x) dx \right)^{\frac{1+\kappa}{\alpha \kappa}}. \tag{13}$$

The coupled entropy $S^C$ is related to the normalized Tsallis entropy $S^{NT}$ and the Tsallis entropy $S^T$ by

$$S_{\alpha,\kappa}^{C} = \frac{S_{\alpha,\kappa}^{NT}}{1+\kappa} = \frac{S_{\alpha,\kappa}^{T}}{(1+\kappa)\int_{x \in X} f^{1+\frac{\alpha \kappa}{1+\kappa}}(x) dx},$$

$$S_{\alpha,\kappa}^{NT}(f(x)) = \frac{S_{\alpha,\kappa}^{T}(f(x))}{\int_{x \in X} f^{1+\frac{\alpha \kappa}{1+\kappa}}(x) dx} = \frac{(1+\kappa) \ln_{\kappa} \left( \int_{x \in X} f^{1+\frac{\alpha \kappa}{1+\kappa}}(x) dx \right)^{\frac{1}{\kappa}}}{\int_{x \in X} f^{1+\frac{\alpha \kappa}{1+\kappa}}(x) dx}. \tag{14}$$

The coupled entropy was shown to increase with the coupling $\kappa$, but at a slower rate than the normalized Tsallis entropy which has been shown to be unstable [31–33]. This contrasts with the Tsallis entropy which decays with the coupling.

Figure 1 shows the coupled Gaussian distributions defined in with $\alpha = 2$ and $Z = \frac{\sigma}{\sqrt{\kappa}} \text{Beta}\left(\frac{1}{2\kappa}, \frac{1}{2}\right)$. Importantly, the value of the density at the mean plus the scale $f(\mu+\sigma)$ is equal to the average density of the distribution defined by (13). The proof of this relationship developed in [6] is summarized here.

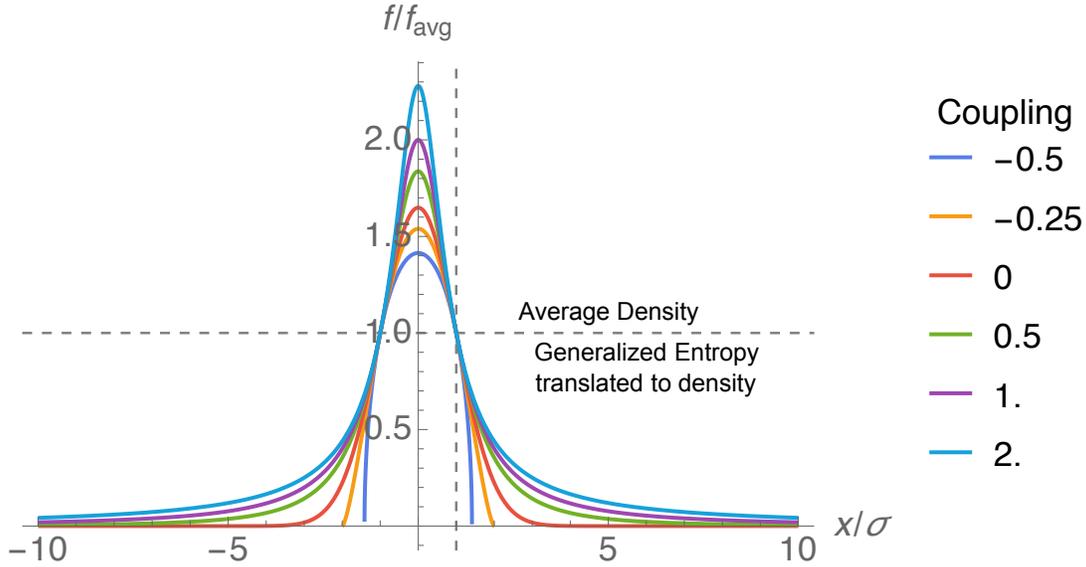

**FIGURE 1: COUPLED GAUSSIAN DISTRIBUTIONS** The coupled Guassians maximizes the Tsallis and Renyi entropies for a constrained generalization of the standard deviation, σ. The density at the width, $f(\mu+\sigma)$ is the value of the generalized entropy transformed to the density domain, thus defining an average density of the distributions. The y axis is scaled by the average density and the x axis is scaled by the scale in order to highlight the symmetry..

*Lemma 1:* The average density, defined in (13) as $f_{\kappa avg} \equiv \left(\int_{x \in X} (f(x))^{1+\frac{2\kappa}{1+\kappa}} dx\right)^{\frac{1+\kappa}{2\kappa}}$, of the heavy-tailed coupled Gaussian distribution $(\kappa > 0, \alpha = 2)$ is equal to the density of the distribution evaluated at $x = \mu + \sigma$.

*Proof:* Without loss of generality we set $\mu = 0$. The proof of this lemma follows from the following calculation,

$$\begin{aligned}
f_{\kappa avg} &= \left( \left( \frac{\sqrt{\kappa}}{\sigma \, \text{Beta}\left(\frac{1}{2\kappa}, \frac{1}{2}\right)} \right)^{\frac{1+3\kappa}{1+\kappa}} \int_{-\infty}^{\infty} \left( \left(1 + \kappa \frac{x^2}{\sigma^2}\right)^{\frac{1+\kappa}{-2\kappa}} \right)^{\frac{1+3\kappa}{1+\kappa}} dx \right)^{\frac{1+\kappa}{2\kappa}} \\
&= \left( \frac{\sqrt{\kappa}}{\sigma \, \text{Beta}\left(\frac{1}{2\kappa}, \frac{1}{2}\right)} \right)^{\frac{1+3\kappa}{2\kappa}} \left( \int_{-\infty}^{\infty} \left(1 + \kappa \frac{x^2}{\sigma^2}\right)^{\frac{1+3\kappa}{-2\kappa}} dx \right)^{\frac{1+\kappa}{2\kappa}} \\
&= \left( \frac{\sqrt{\kappa}}{\sigma \, \text{Beta}\left(\frac{1}{2\kappa}, \frac{1}{2}\right)} \right)^{\frac{1+3\kappa}{2\kappa}} \left( \frac{\sigma \, \frac{\frac{1}{2\kappa}}{\frac{1}{2\kappa}+\frac{1}{2}} \text{Beta}\left(\frac{1}{2\kappa}, \frac{1}{2}\right)}{\sqrt{\kappa}} \right)^{\frac{1+\kappa}{2\kappa}} \\
&= \frac{\sqrt{\kappa}}{\sigma \, \text{Beta}\left(\frac{1}{2\kappa}, \frac{1}{2}\right)} (1+\kappa)^{\frac{1+\kappa}{-2\kappa}} = f(\mu + \sigma) \square
\end{aligned}$$

(15)

The relationship between the generalized mean of the density of the coupled Gaussian and the density at the mean plus the scale motivates an investigation as to whether the generalized mean of random variables drawn from a coupled Gaussian distribution can be used to estimate the scale.

### 3. Estimating the scale parameter of a coupled Gaussian distribution

Equation (11) does not lead to a closed form statistic for the scale. The reason is that the full distribution of the random variable must be estimated in order to calculate the escort probability. In contrast, the sample mean and variance depend on one over the number of samples being an appropriate approximation for the probability of each sample. For the exponential family of distributions the sample mean and standard deviation are sufficient statistics as defined in (2), meaning that all the information from the samples about the distribution's mean and standard deviation is contained in the statistic. The coupled exponential distributions, which are a generalization of the exponential family, have power-law decay for $\kappa \neq 0$ and sufficient statistics may not exist, since these distributions do not satisfy the factorability requirement defined in (3). Estimation of the parameters of a coupled exponential distribution is further complicated by the challenges of estimating the coupling or tail decay. None of the methods for estimating the tail decay such as the Hill estimator and Pickands estimator, which both utilize order statistics, can be said to be uniquely efficient from the perspective of asymptotic variance [34]. Estimators have been developed utilizing the generalized algebra of nonextensive statistical mechanics [35–38]; however, these methods, including a generalization of the maximum likelihood estimate, depend on estimating the escort distribution, precluding a closed form statistic.

In the previous section we established a relationship between the average density and the density at the mean plus the scale. This relationship utilized the generalized mean of the density, which motivates an investigation regarding the generalized mean of the states of the distribution, also known as the fractional moments [39,40]. Lemma 2 establishes the relationship between the scale of coupled Gaussian distributions and the fractional moments of these distributions.

*Lemma 2*: Given a heavy tailed coupled Gaussian distribution with parameter $\kappa$ its scaling parameter $\sigma$ satisfies the following equation

$$\sigma = \begin{cases} \sqrt{\kappa} \left( \int_{-\infty}^{\infty} |x-\mu|^{\frac{1-\kappa}{\kappa}} f(x;\mu,\sigma,\kappa) dx \right)^{\frac{\kappa}{1-\kappa}} & \kappa > 0 \text{ and } \kappa \neq 1 \\ \exp\left( \int_{-\infty}^{\infty} f(x;\mu,\sigma,1) \ln|x-\mu| dx \right) & \kappa = 1. \end{cases} \quad (16)$$

*Proof*: Replacing $f$ with the heavy-tail coupled Gaussian distribution which is (6) with $(\kappa > 0, \alpha = 2)$ and completing the integral shows that the solution is the scale of the distribution. Substituting $x' = x - \mu$, and moving $1/Z$ outside the integral, the integral for $\kappa \neq 1$ is

$$\int_{-\infty}^{\infty} \left( (x')^2 \right)^{\frac{1-\kappa}{2\kappa}} \left( 1 + \kappa \left( \frac{x'}{\sigma} \right)^2 \right)_{+}^{\frac{-1-\kappa}{2\kappa}} dx' = \left( \frac{\sigma}{\sqrt{\kappa}} \right)^{\frac{1}{\kappa}} \text{Beta}\left( \frac{1}{2\kappa}, \frac{1}{2} \right). \quad (17)$$

Note that $1 + \kappa \left( \frac{x'}{\sigma} \right)^2 > 0$ for $\kappa > 0$. Substituting this solution into (16) and simplifying completes the proof for $\kappa \neq 1$

$$\sqrt{\kappa} \left( \left( \frac{\sqrt{\kappa}}{\sigma \text{Beta}\left(\frac{1}{2\kappa},\frac{1}{2}\right)} \right) \left( \frac{\sigma}{\sqrt{\kappa}} \right)^{\frac{1}{\kappa}} \text{Beta}\left(\frac{1}{2\kappa},\frac{1}{2}\right) \right)^{\frac{\kappa}{1-\kappa}} = \sqrt{\kappa} \left( \left( \frac{\sigma}{\sqrt{\kappa}} \right)^{\frac{1-\kappa}{\kappa}} \right)^{\frac{\kappa}{1-\kappa}} = \sigma. \quad (18)$$

For $\kappa = 1$ the distribution is the Cauchy distribution and the generalized mean converges to the geometric mean or equivalently the exponential of the logarithmic average

$$\lim_{\kappa \to 1} \left( \sqrt{\kappa} \left( \int_{-\infty}^{\infty} |x-\mu|^{\frac{1-\kappa}{\kappa}} f(x;\mu,\sigma,\kappa) dx \right)^{\frac{\kappa}{1-\kappa}} \right) = \sqrt{\kappa} \exp\left( \int_{-\infty}^{\infty} \ln|x-\mu| f(x;\mu,\sigma,\kappa) dx \right). \quad (19)$$

Similar to the general case, substituting the Cauchy density for $f$ and calculating the integral the proof is completed. We used *Mathematica* to calculate the integrals in the proof. □

Given the relationship in Lemma 2 we explore the properties of the following empirical estimator for the scale parameter

$$\hat{\sigma} = \begin{cases} \sqrt{\kappa}\left(\dfrac{1}{N}\sum_{i=1}^{N}|x_i - \mu|^{\frac{1-\kappa}{\kappa}}\right)^{\frac{\kappa}{1-\kappa}} & \kappa \neq 1 \\ \exp\left(\dfrac{1}{N}\sum_{i=1}^{N}\ln|x_i - \mu|\right) & \kappa = 1. \end{cases} \qquad (20)$$

Figure 2a shows the case of the Cauchy distribution ($\kappa = 1$) in which case the right side of (20) reduces to the geometric mean. The convergence of the estimator for large sample size is comparable to that of the mean-square error for estimating the standard deviation of the Gaussian distribution. Unfortunately, the convergence rate degrades severely for coupling values different from 1. Figure 2b and c show estimation of the coupled deviation for $\kappa = 0.25$ and $\kappa = 16$ respectively. For $\kappa < 1$ the estimate of the scale parameter is biased low with some anomalous high values leading in some cases to dramatic over-estimation. In contrast, for $\kappa > 1$ the estimate is biased low with some anomalous low values leading to similar under-estimation.

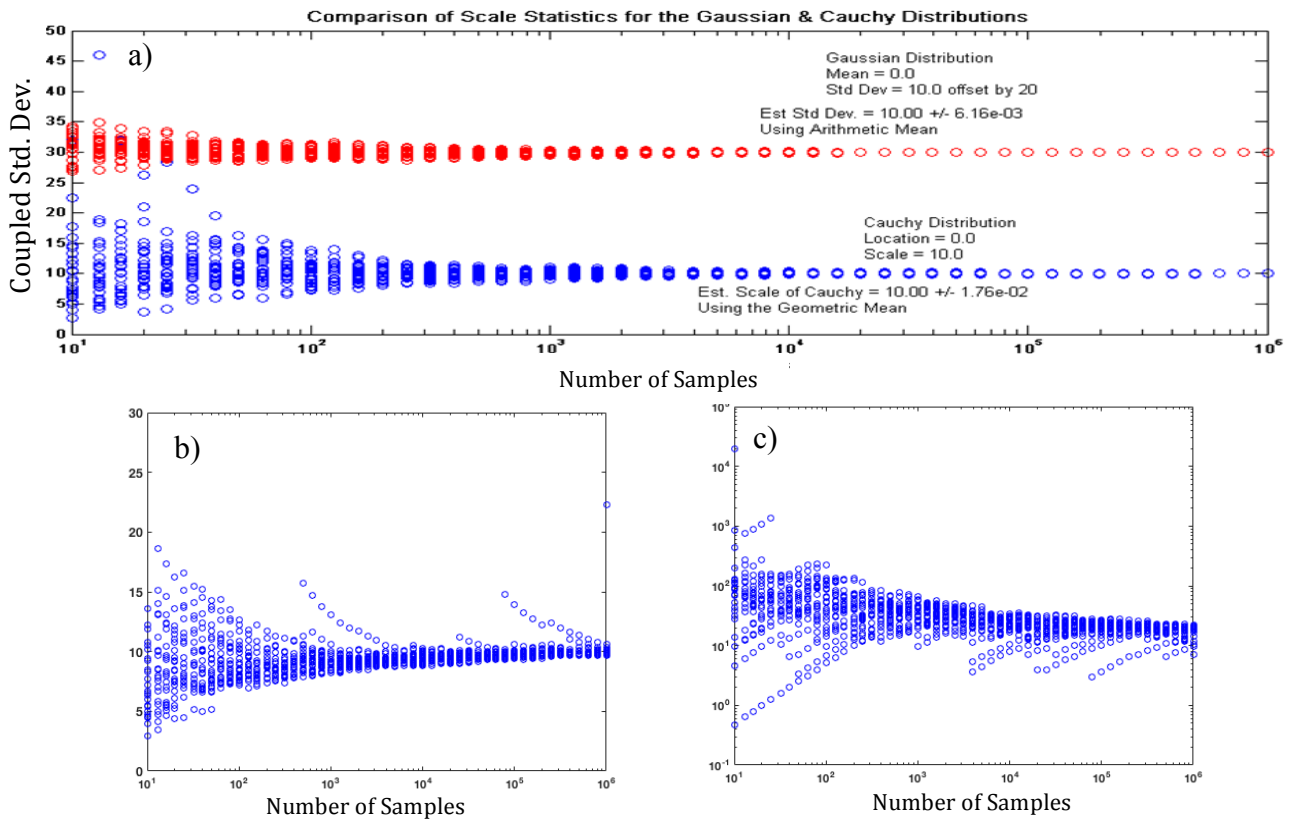

**FIGURE 2 SCALE ESTIMATION WITH THE GENERALIZED MEAN** a) Estimation of the scale for the Cauchy Distribution (κ=1) is with the Geometric Mean (0) from 10 to 1 Million samples. It's convergence is similar to estimation of the variance of the Gaussian distribution using the arithmetic mean. b) Estimation of the scale for the κ=0.25 Coupled Gaussian using the 1.5 generalized mean suffers from anomalous overestimation. c) Estimation of the scale for the κ=16 Coupled Gaussian using the 1.5 generalized mean suffers from underestimation.

The numerical observation of diminishing variance as the sample size increases for the estimate of the scale of the Cauchy distribution leads to the question of whether the geometric mean can be used more generally as a statistic for the coupled Gaussians. Lemma 3 establishes an analytical relationship between the geometric mean and the scale. The lemma makes use of the harmonic number which is defined as

$$H_a \equiv \int_0^1 \frac{1-x^a}{1-x} dx, a > -1, \text{ and} \tag{21}$$

$$H_n \equiv \sum_{k=1}^n \frac{1}{k} \text{ for integers.}$$

The harmonic number arises from the properties of the integral of the logarithm of the gamma function and is approximately ln(a) plus the Euler–Mascheroni constant $\gamma = 0.577$. The Beta function, which is part of the normalization of the coupled Gaussian, is a ratio of gamma functions.

*Lemma 3*: Given a coupled Gaussian distribution $f(x;\mu,\sigma,\kappa) = \frac{1}{Z}\left(1+\kappa\left(\frac{x-\mu}{\sigma}\right)^2\right)_+^{\frac{-1}{2}\left(\frac{1}{\kappa}+1\right)}$ with coupling $\kappa > 0$ the scale $\sigma$ of the distribution is a function of the geometric mean, coupling and a harmonic number,

$$\sigma = 2\sqrt{\kappa}\exp\left(\tfrac{1}{2}H_{\left(-1+\frac{1}{2\kappa}\right)} + \int_{-\infty}^{\infty}\ln|x-\mu|f(x;\mu,\sigma,\kappa)dx\right). \tag{22}$$

*Proof*: Substituting $x' = x - \mu$, the logarithmic average of the coupled Gaussian with $\kappa > 0$ is

$$\frac{1}{2}\left(\frac{\sqrt{\kappa}}{\sigma\,\text{Beta}\left(\frac{1}{2\kappa},\frac{1}{2}\right)}\right)\int_{-\infty}^{\infty}\ln|x'|\left(1+\kappa\left(\frac{x'}{\sigma}\right)^2\right)_+^{\frac{-1}{2}\left(\frac{1}{\kappa}+1\right)}dx'$$

$$= \frac{x}{2}\left(\frac{\sqrt{\kappa}}{\sigma\,\text{Beta}\left(\frac{1}{2\kappa},\frac{1}{2}\right)}\right)\left[{}_2F_1\left(\tfrac{1}{2},\tfrac{1+\kappa}{2\kappa},\tfrac{3}{2},\tfrac{-x^2\kappa}{\sigma^2}\right)\ln|x'| - 2\,{}_pF_q\left(\{\tfrac{1}{2},\tfrac{1}{2},\tfrac{1}{2}+\tfrac{1}{2\kappa}\},\{\tfrac{3}{2},\tfrac{3}{2}\},\tfrac{-x'^2\kappa}{\sigma^2}\right)\right]_{-\infty}^{\infty} \tag{23}$$

$$= \frac{1}{2}\left(\frac{\sqrt{\kappa}}{\sigma\,\text{Beta}\left(\frac{1}{2\kappa},\frac{1}{2}\right)}\right)\left[\left(\frac{-\sigma\text{Beta}\left(\frac{1}{2\kappa},\frac{1}{2}\right)}{\sqrt{\kappa}}\right)\left(H_{\frac{1}{2\kappa}-1}+\ln\left(\frac{4\kappa}{\sigma^2}\right)\right)\right] = \ln\left(\frac{\sigma}{2\sqrt{\kappa}}\right) - \tfrac{1}{2}H_{\frac{1}{2\kappa}-1}.$$

Taking the exponential of this solution and rearranging terms gives (22) completing the proof. For the Cauchy distribution $(\kappa = 1)$ the harmonic number is $H_{-\frac{1}{2}} = -\ln 4$, which simplifies (23) to $\ln\left(\frac{2\sigma}{\sqrt{\kappa}}\right)$ and (22) to $\sigma = \exp\left(\int_{-\infty}^{\infty}\ln|x-\mu|f(x;\mu,\sigma,1)dx\right)$. □

Thus another potential estimator for the scale is

$$\hat{\sigma}_* = 2\sqrt{\kappa}\exp\left(\tfrac{1}{2}H_{\left(-1+\frac{1}{2\kappa}\right)}\right)\prod_{i=1}^N |x_i - \mu|^{1/N}. \tag{24}$$

Figure 3 shows empirical convergence without the anomalies evident for the generalized mean estimator. For plot a $(\kappa = 0.1)$ and b $(\kappa = 0.5)$, 50 estimates given 10 to 1 million samples are plotted on a linear-log scale. Like the estimate for the Cauchy distribution, the estimate for the coupled standard deviation in these cases is unbiased and has a variance on the order of $10^{-2}$ with 1 million samples. While a log-log plot is used for plot c $(\kappa=2)$, due to a large deviation with just 10 samples, the convergence with a large number of samples is still only modestly noisy. It is not until $\kappa>5$, that the affects of the extreme heavy-tail distribution begin to make the estimate less reliable. For $\kappa=10$ (plot d) the convergence is slower but still unbiased. Figure 4 shows the mean and standard deviation of the estimated coupled standard deviation with 1 million samples. Below $\kappa<1$ the standard deviation is less than 0.03. Between $1<\kappa<5$ the standard deviation grows to 0.06, but the mean is still approximately 10.

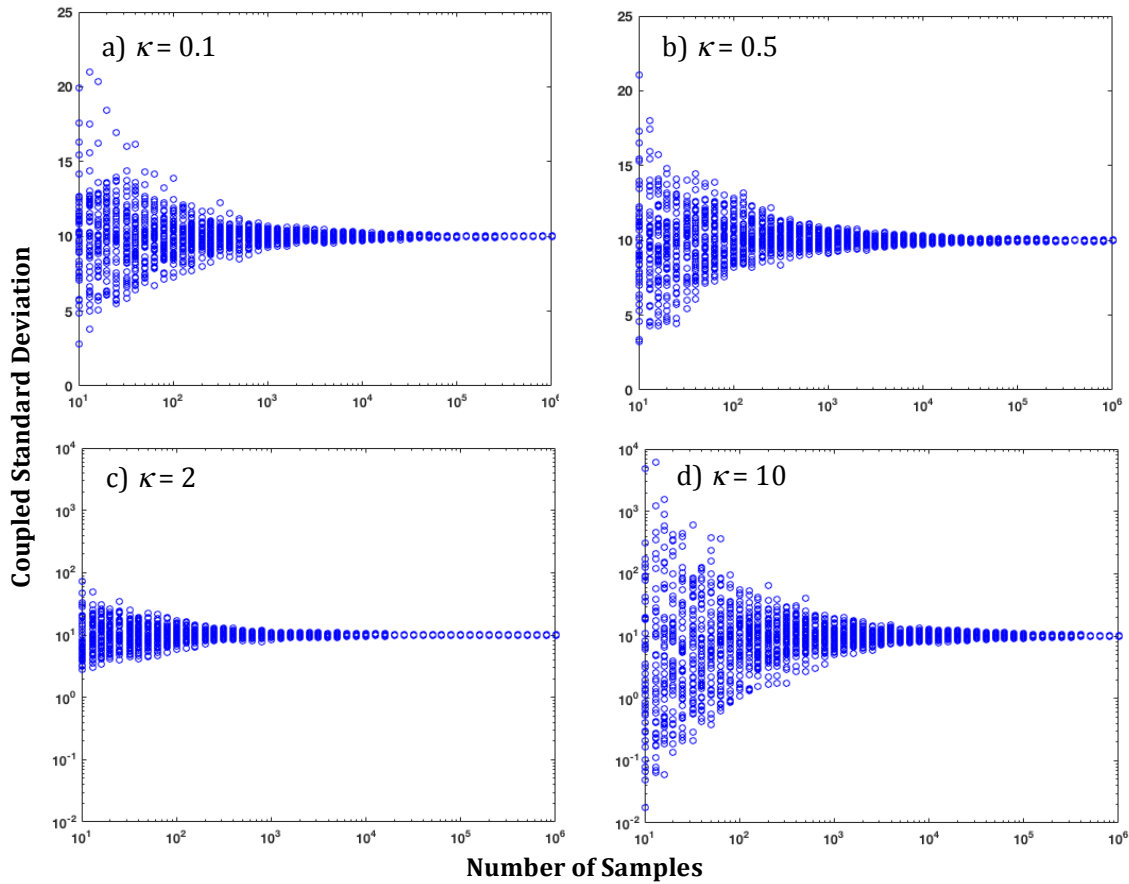

FIGURE 3 SCALE ESTIMATION WITH THE GEOMETRIC MEAN. Scale estimation of the Coupled Gaussian using the geometric mean converges for a wide range of coupling. For coupled Gaussians with κ<1 the estimations are plotted on a linear-log plot; a) κ=0.1 and b) κ=0.5. For coupled Gaussian with the κ>1 a log-log plot is necessary, but the estimate still converges for c) κ=2 and d) κ=10.

…

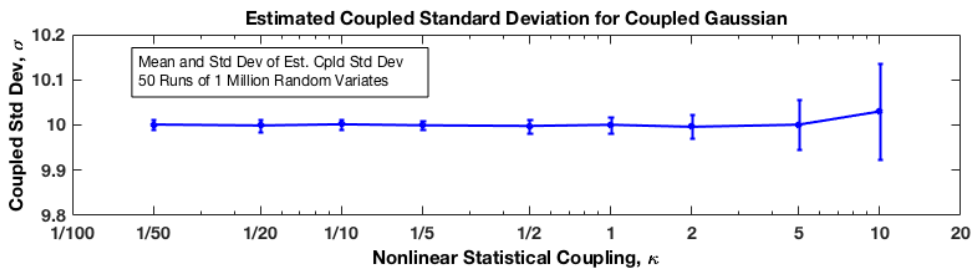

**FIGURE 4** The mean and standard deviation (50 runs of 1 million samples) of the coupled standard deviation (scale) of the coupled Gaussian for coupling parameters ranging from 1/50 to 10.

## 4. Estimating the scale and coupling parameters of coupled Gaussian distribution

Estimating the decay of a heavy-tailed distribution is challenging due to a typically large or infinite source variance. Use of the geometric mean as a statistic provides improved estimated pairs of scale and coupling parameters for coupled Gaussian distributions. Assuming a given geometric mean statistic, selecting a best-fit distribution is reduced from a two to a one-dimensional search. Possible tests of fitness include the maximum likelihood, Kolmogorov-Smirnov and Cramer-Von Mises tests. Shown here are results using the Cramer-Von Mises (CVM) test, which has a high degree of sensitivity and accuracy for coupled Gaussian distributions. A simulated sample set of size one million is used to estimate the underlying scale and coupling parameters. The *p*-values of the CVM tests are shown in Figure 5 for sub-sample sizes of one, ten and one hundred thousand. On the full one million sample set, the test appears to be over-sensitive. The source scale $\sigma$ is 1.00 in each case. For source $\kappa = 1.00$, the estimate is accurate to the second decimal place. Some error occurs for $\kappa = 0.50$ and $\kappa = 2.00$. For $\kappa = 10.0$ the estimation is not as reliable. In this case, limiting to 1000 samples provides a rough approximation, but using more samples causes the search for an optimal pair to fail.

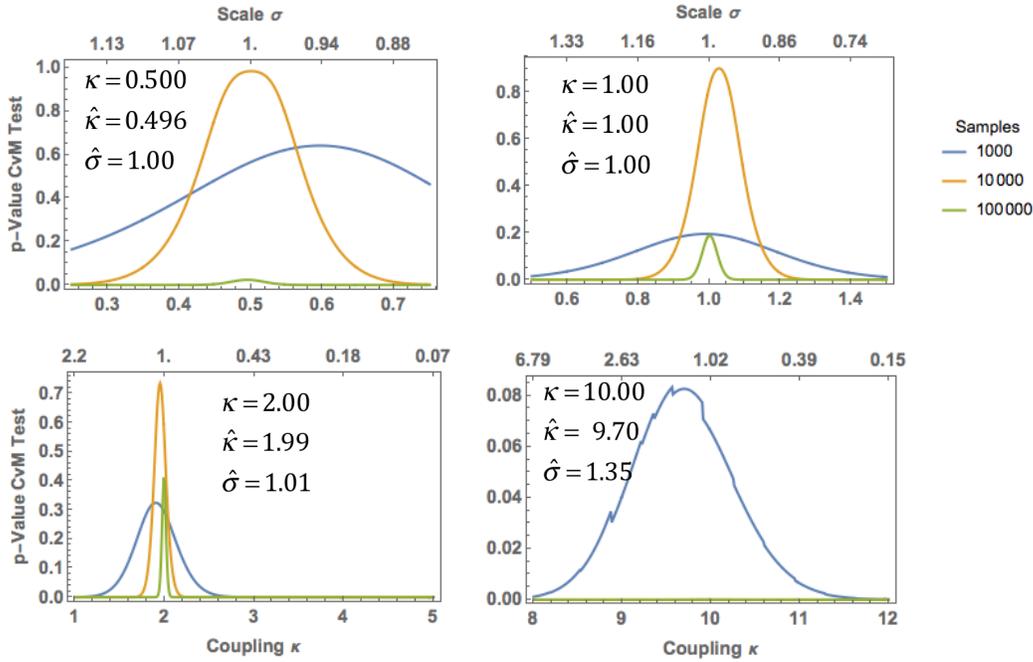

**FIGURE 5** The *p*-value results of the Cramer-Von Mises test for scale-coupling parameter pairs of the coupled Gaussian distribution. The source scale is 1.00 in each case. For source $\kappa=1.00$ the estimate is exact to the second decimal place. Some error occurs for $\kappa=0.50$ and $\kappa=2.00$. For $\kappa=10.0$ the estimation is not as reliable. Limiting to 1000 samples a rough approximation is possible, but using more samples causes the test to fail.

## 5. Conclusion and Discussion

Complex systems influenced by nonlinearities can present challenges to proper characterizations of the uncertainties of underlying process. The basic assumptions on the estimated variance and entropy as fundamental characterizations of uncertainty are not always reliable when the system is highly nonlinear. One effect of nonlinearities, as we have discussed, is the appearance of heavy-tailed statistical distributions. A two-parameter family of distributions, referred to as coupled exponential distributions, is used to provide useful statistical parameterizations for nonlinear systems. These distributions maximize the Tsallis, Renyi and coupled-entropies, and include the generalized Pareto $(\alpha=1, \kappa: \text{shape})$, the Student's *t* $(\alpha=2, \kappa: \text{inverse of degree of freedom})$, Cauchy $(\alpha=2, \kappa=1)$, exponential $(\alpha=1, \kappa=0)$ and Gaussian $(\alpha=2, \kappa=0)$ distributions.

Analytically the generalized mean of the $\frac{1-\kappa}{\kappa}$ power of a coupled Gaussian distribution is equal to the scale parameter divided by the shape parameter $\sigma/\kappa$. However, our numerical experiments show that only the geometric mean ($\kappa=1$) for the Cauchy distribution is an unbiased estimator apparently free from anomalies. We have shown that the geometric mean can be used as a statistic for other members of the coupled Gaussian family if the estimator incorporates the harmonic number. The estimator for the scale in coupled Gaussians is provided in (24). Simulated convergence properties of the estimator are shown in Figure 3 and show that for $\kappa < 5$, estimates with minimal bias and variance appear to be possible.

This numerical evidence for the geometric mean as a statistic for the scale of coupled exponentials suggests further theoretical investigation is merited. In particular, we conjecture that $\hat{\sigma}$ defined in (24) is an efficient estimator for the scale of the coupled Gaussian distribution, meaning that it minimizes mean square error. If so, this opens the possibility of generalizing the factorization theorem [9] for sufficient statistics. In [10] the factorization of multivariate coupled exponential distributions using a generalization of the product function has been described. It is possible that the coupled exponential family can be defined more formally [11] in terms of a set of natural parameters and sufficient statistics using generalizations of the product function.

## Acknowledgements

The symbolic computational capabilities of *Mathematica* were instrumental in completing the integral for the logarithmic average of the coupled Gaussian distributions, which lead to the discovery the proposed estimator for the scale.